\documentstyle[preprint,aps]{revtex}
\tightenlines                                                                                                                                                                
\newcommand{\be}{\begin{equation}}
\newcommand{\ee}{\end{equation}}
\newcommand{\bea}{\begin{eqnarray}}
\newcommand{\eea}{\end{eqnarray}}
\newcommand{\ba}{\begin{array}}
\newcommand{\ea}{\end{array}}

\newcommand{\th}{\theta}

\newcommand{\la}{\lambda}
\newcommand{\La}{\Lambda}
\newcommand{\de}{\delta}

\newcommand{\pa}{\partial}

\newcommand{\no}{\nonumber}

\newcommand{\Om}{\Omega}
\newcommand{\om}{\omega}

\newcommand{\sres}{\mbox{sres}}
\newcommand{\Str}{\mbox{Str}}

\newcommand{\lb}{\label}
\newcommand{\na}{\nabla}

\begin{document}

\title{On the Miura and B\"acklund transformations associated\\
with the supersymmetric Gelfand-Dickey bracket}

\author{Jiin-Chang Shaw$^1$ and Ming-Hsien Tu$^2$ }
\address{
$^1$ Department of Applied Mathematics, National Chiao Tung University, \\
Hsinchu, Taiwan, \\
$^2$ Department of Physics, National Tsing Hua University, \\
Hsinchu, Taiwan
}
\date{\today}
\maketitle

\begin{abstract}
The supersymmetric version of the Miura and B\"acklund transformations 
associated with the supersymmetric Gelfand-Dickey bracket are investigated
from the point of view of the Kupershmidt-Wilson theorem.
\end{abstract}

\newpage
Soon after the work of Zamolochikov \cite{Z}, the $W$-algebras 
have been paid much attention in two-dimensional conformal field theory 
and integrable systems \cite{BS}.
Now it is well known that the classical version of the $W_n$-algebras arise
naturally as the second Gelfand-Dickey (GD) bracket of the $n$-th generalized
Korteweg-de Vries (KdV) hierarchy \cite{DS,Dic} defined by the Lax operator 
$L=\pa^n+u_{n-1}\pa^{n-1}+\cdots+u_0$. In such connection, 
the Kupershmidt-Wilson (KW) theorem\cite{KW,D2} play an important role 
in such a way that under the factorization $L=(\pa-\phi_n)(\pa-\phi_{n-1})\cdots(\pa-\phi_1)$
the second GD bracket is transformed into a much simple 
one in terms of the Miura fields $\phi_i$. 
Thus the factorization not only provides the Miura transformation
which maps the $n$-th KdV hierarchy to the corresponding modified hierarchies but
also gives a free-field realization of the associated classical $W_n$-algebras \cite{D3}. 
On the other hand, as observed by Adler in \cite{Ad} that the cyclic permutation of the 
Miura fields $\phi_i$ in the factorized Lax operator  generates
the B\"acklund transformation of the $n$-th KdV hierarchy. 
Hence from the point of view of the KW theorem, one has a unified description
of the Miura and B\"acklund transformations associated with the second
GD structure for the $n$-th KdV hierarchy. In general, the above scheme 
is encoded in a particular form of the Lax operator and its associated 
Poisson structure. Several integrable systems have been studied based on this scheme. 
For example, the KW theorem and Adler's work have been generalized to the constrained KP 
hierarchy \cite{Y,ANP,D4,MR,C,AFGZ} which is defined by the pseudo-differential 
Lax operator of the form 
$L=\pa^n+u_{n-1}\pa^{n-1}+\cdots+u_0+\sum_{i=1}^m q_i\pa^{-1}r_i$.

Recently, the supersymmetric second GD bracket has been 
constructed \cite{FR1,FR2,FR3} for the generalized supersymmetric KdV(SKdV) 
hierarchy. In particular, the supersymmetric KW theorem which enables one 
to reduce the supersymmetric second GD bracket to a vastly simple form via a 
supersymmetric Miura transformation has been investigated \cite{FR1,FS}. 
However, in contrast to the bosonic counterpart,there are some issues 
associated with the supersymmetric GD bracket still unexplored. 
In this letter we consider two of them: the first is the factorization 
property of the supersymmetric GD bracket defined by the Lax operator 
which can be expressed as a ratio of two purely superdifferential 
operator $L=AB^{-1}$. The other is to generalize Alder's work to the 
supersymmetric GD bracket. Once these goals can be achieved,
we will have a unified way to describe the Miura and B\"acklund transformations 
associated with the supersymmetric GD bracket, which, to the best of our knowledge, 
has not been reported in the literature.  

To begin with, we consider the superdifferential Lax operator of the form
\be
L=D^n+U_{n-1}D^{n-1}+\cdots+U_0
\lb{lax1}
\ee
where the supercovariant derivative $D\equiv\pa_{\th}+\th\pa$ satisfies 
$D^2=\pa$, $\th$ is the Grassmann variable ($\th^2=0$) which together
with the even variable $x\equiv t_1$, defines the $(1|1)$ superspace with
coordinate $(x,\th)$. The coefficient $U_i$ are superfields that depend on
the variables $x$, $\th$, $t_i$ and can be represented by $U_i=u_i(t)+\th v_i(t)$.
Since $L$ is assumed to be homogeneous under $Z_2$-grading and hence $|U_i|=n+i$
(mod 2). We will introduce the Poisson bracket associated withthe 
Lax operator on functionals of the form:
\be
F(U)=\int_B f(U)
\ee
where $f(U)$ is a homogeneous differential polynomialof the 
$U_i$ and $\int_B\equiv \int dxd\th$ is the Berezin integral 
such thatif $f(U)=a(u,v)+\th b(u,v)$ then $\int_Bf(U)=\int b$. 
The supercovariant derivative $D$ satisfies the 
supersymmetric version of the Leibniz rule \cite{Manin}:
\be
D^iU=\sum_{k=0}^{\infty}(-1)^{|U|(i-k)} {i\brack k}U^{[k]}D^{i-k}
\label{leib}
\ee
where $U^{[k]}\equiv (D^kU)$ and the super-binomial 
coefficients ${i\brack k}$ are defined by
\be
{i\brack k}=
\left\{
\ba{l}
{[i/2]\choose [k/2]} \qquad \mbox{for $ 0\leq k\leq i$ and $(i,k)\ne (0,1)$ mod 2}\\
(-1)^{[k/2]}{-i+k-1\brack k}\qquad \mbox{for $i<0$}\\
0\qquad \mbox{otherwise}
\ea
\right.
\ee
For a given super pseudo-differential operator $P=\sum p_iD^i$ we can 
define its super-residue as $\sres P=p_{-1}$ and its supertrace 
as $\Str P=\int_B\sres P$.It can be shown that, for any two super 
pseudo-differential operators $P$ and $Q$,$\Str [P,Q]=0$ for 
$[P,Q]\equiv PQ-(-1)^{|P||Q|}QP$ and hence
$\Str PQ=(-1)^{|P||Q|}\Str QP$.

Finally, given a functional $F(U)=\int_Bf(U)$, we define its gradient $d_LF$ by
\be
d_LF=\sum_{k=0}^{n-1}(-1)^kD^{-k-1}\frac{\de f}{\de U_k}
\lb{grad}
\ee
where the variational derivative is defined by
\be
\frac{\de f}{\de U_k}=\sum_{i=0}^{\infty}(-1)^{|U_k|i+i(i+1)/2}
D^i\frac{\pa f}{\pa U_k^{[i]}}
\lb{vari}
\ee
The supersymmetric GD bracket defined by the Lax operator (\ref{lax1})
is given by \cite{FR1,FR2,FR3}
\be
\{F,G\}(L)=(-1)^{|F|+|G|+|L|}\Str[L(d_LFL)_+d_LG-(Ld_LF)_+Ld_LG]
\lb{sgd}
\ee
where $()_+$ denotes the differential part of a super pseudo-differential operator.
It has been shown \cite{FR1,FR3} that (\ref{sgd}) indeed defines a Hamiltonian 
structure, namely it is antisymmetric and satisfies the super Jacobi identity.
From (\ref{sgd}) the fundamental Poisson brackets of the $U_k$ can be read off
by taking $F$ and $G$ are linear functionals.
The bracket (\ref{sgd}) has a very nice factorization property
associated with the Lax operator (\ref{lax1}). If $L=AB$ where
both $A$ and $B$ are superdifferential operators, then it can be proved \cite{FS}that 
\be
\{F,G\}(L)=\{F,G\}(B)+(-1)^{|B|}\{F,G\}(A)
\lb{prod}
\ee
Note that there is a relative sign $(-1)^{|B|}$ 
between the two terms, which
disappears in the bosonic counterpart. Now let us factorize 
$L=(D-\Phi_n)(D-\Phi_{n-1})\cdots(D-\Phi_1)$, 
then such supersymmetric 
Miura transformation defines the $U_i$ 
as polynomials in the Miura fields $\Phi_i$. 
Since the supersymmetric GD bracket for the operator
$D-\Phi_i$ can be easily calculated as 
$\{F,G\}(D-\Phi_i)=-\int_B(D\frac{\de f}
{\de \Phi_i})\frac{\de g}{\de \Phi_i}$.
Hence under the factorization the second bracket becomes
\be
\{F,G\}(L)=\int_B\sum_{i=1}^n(-1)^i(D\frac{\de f}
{\de \Phi_i})\frac{\de g}{\de \Phi_i}
\ee
which implies that the fundamental brackets of the Miura fields
$\Phi_i$ are given by \cite{FR1,FR2}
\be
\{\Phi_i(X),\Phi_j(Y)\}=(-1)^i\de_{ij}D\de(X-Y)
\lb{free}
\ee
where $X=(x,\th)$, $Y=(y, \om)$ and $\de(X-Y)\equiv\de(x-y)(\th-\om)$. 
This result is what we called the supersymmetric KW theorem.
Eq.(\ref{free}) also enables us to write down 
the fundamental brackets of the $U_k$
through the super Miura transformation. 
Notice that the alternating
signs in (\ref{free}) are crucial since it is only 
this choice which will lead to a 
closed Poisson algebra among the $U_k$.

In fact, it has been shown \cite{Yu,DH,Yung} 
that the supersymmetric second GD
bracket (\ref{sgd}) still defines a Hamiltonian 
structure even when the super Lax operator
are homogeneous super pseudo-differential 
operators on the supermanifold of the form
\be
\La_n=D^n+V_{n-1}D^{n-1}+\cdots+V_0+V_{-1}D^{-1}+\cdots\qquad n\in Z
\lb{lax2}
\ee
In other words, the fundamental Poisson brackets of the $V_i$ can be calculated from
the supersymmetric GD bracket defined by $\La_n$. 
However, motivated by the recent studies on 
super pseudo-differential operators with reductions and truncations \cite{BD,AR,AD} 
it would be interesting to consider the supersymmetric GD bracket defined by 
the Lax operator $L=AB^{-1}$ where both $A$ and $B$ are superdifferential operators; 
in fact, in view of (\ref{prod}), this situation is reduced to consider the case $L=B^{-1}$. 
More precisely, if $B=\sum B_iD^i$ be an invertible superdifferential Lax operator
such that the Poisson brackets among the $B_i$ be given by (\ref{sgd}).
Then taking $L=B^{-1}=\sum_jU_jD^j$ we can  ask what the 
Poisson bracket associated with $L$ is. To answer this question,  suppose 
$F$ is a functional of $L$, then the infinitesimal change of $F$ gives
\bea
\de F&=&(-1)^{|F|+|L|+1}\Str(\de Ld_LF)\no\\
&=&(-1)^{|F|+|B|+1}\Str(\de Bd_BF)
\eea
which implies
\be
d_BF=(-1)^{|F||L|+|L|+1}Ld_LFL
\lb{rela1}
\ee
Substitution (\ref{rela1}) into (\ref{sgd}) we find
\bea
\{F,G\}(B)&=&(-1)^{|F|+|G|+|B|}\Str[B(d_BFB)_+d_BG-(Bd_BF)_+Bd_BG]\no\\
&=&(-1)^{|F|+|G|+1}\Str[L(d_LFL)_+d_LG-(Ld_LF)_+Ld_LG]\no\\
&=&(-1)^{|L|+1}\{F,G\}(L)
\lb{inv}
\eea
Now using (\ref{prod}) and (\ref{inv}), the supersymmetric GD bracket 
under the factorization $L=AB^{-1}$ can be easily derived as
\bea
\{F,G\}(L)&=&\{F,G\}(B^{-1})+(-1)^{|B|}\{F,G\}(A)\no\\
&=&(-1)^{|B|}(\{F,G\}(A)-\{F,G\}(B))
\lb{quot}
\eea
Again, up to a phase factor, this result is similar to the bosonic case.
Furthermore, if both $A$ and $B$ are factorized into the multiplication form 
(generalized Miura transformation) such that
\be
L=(D-\Phi_n)\cdots(D-\Phi_1)(D-\Psi_1)^{-1}\cdots(D-\Psi_m)^{-1},
\ee
then the only nonzero Poisson brackets among the
generalized Miura fields $\Phi_i$ and $\Psi_i$ are given by
\bea
\{\Phi_i(X),\Phi_j(Y)\}&=&(-1)^{m+i}\de_{ij}D\de(X-Y)\no\\
\{\Psi_i(X),\Psi_j(Y)\}&=&(-1)^{m+i+1}\de_{ij}D\de(X-Y)\\
\{\Phi_i(X),\Psi_j(Y)\}&=&0\no
\eea
These brackets would be possible to provide the free-field realizations of  some 
extended classical $W$-superalgebras.

So far, we only discuss the second Hamiltonian structure. To
obtain the first structure, one can deform the second structure
by shifting $L\rightarrow L+\la$ (or $U_0\rightarrow U_0+\la$) and
extract the first structure from the term proportional to $\la$.

In the following, let us go back to the integrable hierarchy defined by the 
supersymmetric Lax operator of the form $L=D^{2l}+\sum_{i=0}^{2l-1}U_iD^i$, 
the so-called generalized $l$-th order super KdV(SKdV) hierarchy\cite{OP,FMR}.
Since it is a reduction of the super pseudo-differential operator $\La_2$ 
in such a way that $L=(\La_2^l)_+$, and hence we also call this hierarchy
the $\mbox{SKP}_2^{(l)}$ hierarchy\cite{OP,FMR}.

The evolution equations of the even order $l$-th SKdV hierarchy are given by
\be
\frac{\pa L}{\pa t_k}=[L^{k/l}_+,L]\qquad (k=1,2,\cdots)
\lb{laxeq}
\ee
Due to the Poisson structure (\ref{sgd}), one can give a Hamiltonian flow description 
of the Lax equation (\ref{laxeq}) as
\be
\frac{\pa L}{\pa t_k}=\{H_k, L\}
\lb{hameq}
\ee
where the Hamiltonians $H_k$ are defined by
\be
H_k=\frac{l}{k}\Str L^{k/l}\equiv \int_B h_k
\lb{hamil}
\ee
which form a infinite number of conservation quantities of the hierarchy.
If we factorize the even order super Lax operator as 
$L=(D-\Phi_{2l})(D-\Phi_{2l-1})\cdots(D-\Phi_1)$ then, by KW theorem, 
the corresponding Miura fields $\Phi_i$ satisfy the modified hierarchy equations
\be
\frac{\pa \Phi_i}{\pa t_k}=\{H_k, \Phi_i\}=(-1)^i(D\frac{\de h_k}{\de \Phi_i})
\lb{modif1}
\ee
In the following, we would like to show that the whole set of equation (\ref{modif1})
is invariant under the cyclic permutation 
$\Om$: $\Phi_1\mapsto \Phi_2$, $\cdots$, $\Phi_{2l}\mapsto \Phi_1$.
To see this, let us denote by $L_{\Om}$ and 
$M_{\Om}^{(k)}\equiv (L_{\Om}^{k/l})_+$ the operators obtained by
the cyclic permutation $\Om$. Namely, 
$L_{\Om}=(D-\Phi_1)(D-\Phi_{2l})\cdots(D-\Phi_2)$. Thus we have
$L_{\Om^j}$ and $M_{\Om^j}^{(k)}$ the operators obtained by the permutation $\Om^j$.  
 It is obvious that $L_{\Om^{2l}}=L$ and $M_{\Om^{2l}}^{(k)}=M^{(k)}$. 
Using these permuted operators, we will show that the modified hierarchy 
equations (\ref{modif1}) can be expressed as
\be
\frac{\pa \Phi_i}{\pa t_k}=\na_iM_{\Om^{i-1}}^{(k)}-M_{\Om^i}^{(k)}\na_i
\lb{modif2}
\ee
where $\na_i\equiv D-\Phi_i$. Evidently, eq.(\ref{modif2}) is consistent 
with the Lax equation (\ref{laxeq}) since
\bea
\frac{\pa L}{\pa t_k}&=&-\sum_{i=1}^{2l}
\na_{2l}\cdots\na_{i+1}\frac{\pa \Phi_i}{\pa t_k}\na_{i-1}\cdots\na_1\no\\
&=&-\sum_{i=1}^{2l}\na_{2l}\cdots\na_{i+1}(\na_iM_{\Om^{i-1}}^{(k)}-
M_{\Om^i}^{(k)}\na_i)\na_{i-1}\cdots\na_1\no\\
&=&[M^{(k)},L]
\eea
On the other hand, (\ref{modif2}) makes sense only if 
the order of the right hand side is zero.
This can be shown by observing the fact that the permuted 
operators $L_{\Om^i}$ obey the recursive relation
\bea
L_{\Om^i}&=&\na_iL_{\Om^{i-1}}\na_i^{-1}\no\\
& &\cdots\no\\
&=&\na_i\cdots\na_1L\na_1^{-1}\cdots\na_i^{-1}
\lb{recur}
\eea
Hence the operator $M_{\Om^i}^{(k)}$ can be written by
\bea
M_{\Om^i}^{(k)}&\equiv&(L_{\Om^i}^{k/l})_+\no\\
&=&(\na_iM_{\Om^{i-1}}^{(k)}\na_i^{-1})_+\no\\
&=&\na_iM_{\Om^{i-1}}^{(k)}\na_i^{-1}-(\na_iM_{\Om^{i-1}}^{(k)}\na_i^{-1})_-
\eea
which leads to
\bea
\na_iM_{\Om^{i-1}}^{(k)}-M_{\Om^i}^{(k)}\na_i
&=&(\na_iM_{\Om^{i-1}}^{(k)}\na_i^{-1})_-\na_i\no\\
&=&(D\chi_i^{-1}M_{\Om^{i-1}}^{(k)}\chi_i)_0
\lb{permut}
\eea
where we have rewritten $\na_i=\chi_iD\chi_i^{-1}$ with 
$\chi_i\equiv \exp(\int^x (D\Phi_i))$ and used the fact
$(AD^{-1})_-=(A)_0D^{-1}+A_-D^{-1}$. 

Now we are ready to show that  (\ref{modif2}) is, in fact, equivalent to the modified 
hierarchy equation (\ref{modif1}). From (\ref{permut}) and (\ref{recur}) we have
\bea
\frac{\pa \Phi_i}{\pa t_k}&=&((\na_iM_{\Om^{i-1}}^{(k)}\na_i^{-1})_-\na_i)_0\no\\
&=&\sres(\na_iM_{\Om^{i-1}}^{(k)}\na_i^{-1})\no\\
&=&\sres(\na_iL_{\Om^{i-1}}^{k/l}\na_i^{-1})-
\sres(\na_i(L_{\Om^{i-1}}^{k/l})_-\na_i^{-1})\no\\
&=&\sres(\na_i\cdots\na_1L^{k/l-1}\na_{2l}\cdots\na_{i+1})+
\sres(\na_{i-1}\cdots\na_1L^{k/l-1}\na_{2l}\cdots\na_i)
\lb{rela2}
\eea
On the other hand, from (\ref{hamil}) and (\ref{modif1}) we have
\bea
(-1)^i(D\frac{\de h_k}{\de \Phi_i})
&=&[\na_i, \sres(\na_{i-1}\cdots\na_1L^{k/l-1}\na_{2l}\cdots\na_{i+1})]\no\\
&=&\sres[\na_i, \na_{i-1}\cdots\na_1L^{k/l-1}\na_{2l}\cdots\na_{i+1}]\no\\
&=&\mbox{r.h.s of (\ref{rela2})}
\eea

Therefore there are two different expressions for the dynamics of the 
Miura fields $\Phi_i$. However it is more clear 
from (\ref{modif2}) rather (\ref{modif1})  that
the whole set of equation  is invariant under 
the cyclic permutation $\Om$ and
hence generates the B\"acklund transformation 
of the hierarchy, the $2l$-th
power of which is just an identical transformation. 
Thus,  from the Lax formalism
point of view, we can identify the one-step permutation 
$\Om: L_{\Om^{i-1}}\rightarrow L_{\Om^i}$ as an 
elementary B\"acklund transformation
due to the fact that the form of the Lax operator 
and the hierarchy equation are preserved
under such transformation.  
Moreover, from (\ref{recur}) such one-step 
permutation can be viewed as a 
gauge transformation triggered by the gauge 
operator $T_i=\chi_i D \chi_i^{-1}$, i.e. ,
$L_{\Om^i}=T_iL_{\Om^{i-1}}T_i^{-1}$ and the 
even function $\chi_i$ satisfies
\be
\frac{\pa \chi_i}{\pa t_k}=\chi_i\int^x(D\frac{\pa \Phi_i}{\pa t_k})_0=
(M_{\Om^{i-1}}^{(k)}\chi_i)_0
\ee
which for the $t_l$-flow implies $(L_{\Om^{i-1}}\chi_i)_0=0$. Hence $\chi_i$ is a 
wavefunction associated with the Lax operator $L_{\Om^{i-1}}$. Similarly, we can
also consider the anti-cyclic permutation $\Om^{-1}: L_{\Om^i}\rightarrow L_{\Om^{i-1}}$ 
where the B\"acklund transformation is triggered by the gauge operator
$S_i=\chi_i D^{-1}\chi_i^{-1}$. It is not hard to show that $\chi_i^{-1}$ is a
adjoint wavefunction associated with the Lax operator $L_{\Om^i}$, namely
it satisfies $\pa \chi_i^{-1}/\pa t_k=-(M_{\Om^i}^{(k)^*}\chi_i^{-1})_0$ 
and $(L_{\Om^i}^{*}\chi_i^{-1})_0=0$. These B\"acklund transformations
have been applied to several super integrable systems to obtain their soliton solutions
\cite{Liu,LM,ST,ANP2}.

Finally, we would like to remark the canonical property of these elementary
B\"acklund transformations.
Since the parity of the gauge operator $T_i$ 
(or $S_i$) is odd thus each of the Hamiltonian defined by $L_{\Om^i}$ in (\ref{hamil})
is equal to the minus one defined by $L_{\Om^{i-1}}$.  This fact
together with the Hamiltonian flow equation (\ref{hameq}) 
concludes that the second bracket defined
by $L_{\Om^i}$ is also the minus one defined by $L_{\Om^{i-1}}$.  
This is indeed the case because
\bea
\{F,G\}(L_{\Om^i})&=&\{F,G\}(\na_iL_{\Om^{i-1}}\na_i^{-1})\no\\
&=&-(\{F,G\}(\na_iL_{\Om^{i-1}})-\{F,G\}(\na_i))\no\\
&=&-\{F,G\}(L_{\Om^{i-1}})
\eea
 
In summary, we have derived the generalized Miura transformation
associated with the supersymmetric GD bracket, which might be
possible to provide the free-field realizations of some extended
classical $W$-superalgebras. We have also investigated the 
B\"acklund transformations of the even order SKdV hierarchy
from the KW theorem point of view. We found
that the cyclic and anti-cyclic permutations of the Miura fields 
correspond to the elementary B\"acklund transformations of the 
hierarchy. The gauge operators and the canonical property of the
elementary  B\"acklund transformations were also revealed quite naturally
via this formulation.

{\bf Acknowledgments\/}
We would like to thank Prof. W. J. Huang for helpful discussions.
This work is supported by the National Science Council of Taiwan 
under grant No. NSC-87-2811-M-007-0025.
 One of us (M.H.T) also wishes to thank Center for Theoretical Sciences 
of National Science Council of Taiwan for partial support.


\begin{thebibliography}{99}
\bibitem[1]{JCS} e-mail address: shaw@math.nctu.edu.tw
\bibitem[2]{MHT} e-mail address: mhtu@phys.nthu.edu.tw

\bibitem{Z}
A. B. Zamolochikov, {\it Theor. Math. Phys.\/} {\bf 65\/}, 1205 (1985).

\bibitem{BS}
P. Bouwknegt  and  K. Schoutens, {\it Phys. Rep.\/} {\bf 223\/}, 183 (1993).
(Singapore:  World Scientific 1995) and references therein.

\bibitem{DS}
V. G. Drinfeld and V. V. Sokolov, {\it J. Sov. Math.\/} {\bf 30\/}, 1975 (1985).

\bibitem{Dic}
L. A. Dickey, {\it  Soliton Equations and Hamiltonian Systems \/}
(Singapore: World Scintific 1991).

\bibitem{KW} 
B. A. Kupershmidt and G. Wilson,  {\it Invent. Math.\/} {\bf 62\/}, 403 (1981).

\bibitem{D2} 
L. A. Dickey, {\it Commun. Math. Phys.\/} {\bf 87\/}, 127 (1983).

\bibitem{D3}
L. A. Dickey, {\it Acta. Appl. Math.\/} {\bf 47\/}, 243 (1997) 
and references therein.

\bibitem{Ad}
M. Adler, {\it Commun. Math. Phys.\/} {\bf 80\/}, 517 (1981).

\bibitem{Y} 
F. Yu,  {\it Lett. Math. Phys.\/} {\bf 29\/}, 175 (1993).

\bibitem{ANP} 
H. Aratyn, E. Nissimov and S. Pacheva, {\it Phys. Lett.\/} 
{\bf B314\/}, 41 (1993)  .

\bibitem{D4} 
L. A. Dickey,  {\it Lett. Math. Phys.\/} {\bf 35\/}, 229 (1995).

\bibitem{MR} 
J. Mas and E. Ramos, {\it Phys. Lett.\/} {\bf B351\/}, 194 (1995).

\bibitem{C} 
Y. Cheng, {\it Commun. Math. Phys.\/} {\bf 171\/}, 661 (1995); 
{\it Lett. Math. Phys.\/} {\bf 33\/}, 159 (1995); 
{\it Lett. Math. Phys.\/} {\bf 36\/}, 35 (1996).

\bibitem{AFGZ}
H. Aratyn, L. A. Ferreira, J. F. Gomes and A. H. Zimerman, 
{\it J. Math. Phys.\/} {\bf 38\/}, 1559 (1997).

\bibitem{FR1}
J. M. Figueroa-O'Farrill and E. Ramos, {\it Phys. Lett.\/} 
{\bf B262\/}, 265 (1991).

\bibitem{FR2}
J. M. Figueroa-O'Farrill and E. Ramos, {\it Nucl. Phys.\/} 
{\bf B368\/}, 361 (1992).

\bibitem{FR3}
J. M. Figueroa-O'Farrill and E. Ramos, {\it Commun. Math. Phys.\/} 
{\bf 145\/}, 43 (1992).

\bibitem{FS}
J. M. Figueroa-O'Farrill and S. Stanciu, {\it Mod. Phys. Lett.\/} 
{\bf A36\/}, 2767 (1995).

\bibitem{Manin}
Y. I. Manin and A. O. Radul, {\it Commun. Math. Phys.\/} 
{\bf 98\/}, 65 (1985).

\bibitem{Yu}
F. Yu, {\it J. Math. Phys.\/} {\bf 33\/}, 3180 (1992).

\bibitem{DH}
A. Das and W. J. Huang, {\it Mod. Phys. Lett.\/} 
{\bf A7\/}, 2159 (1992).

\bibitem{Yung}
C. M. Yung, {\it Mod. Phys. Lett.\/} {\bf A8\/}, 129 (1992).

\bibitem{BD}
J. C. Brunelli and A. Das, {\it Phys. Lett.\/} {\bf B337\/}, 303 (1994); 
{\it Phys. Lett.\/}  {\bf B354\/}, 307 (1995); 
{\it Int. J. Mod. Phys.\/} {\bf A10}, 4563 (1995).

\bibitem{AR}
H. Aratyn and C. Rasinariu,{\it Phys. Lett.\/} {\bf B391\/}, 99 (1997).

\bibitem{AD}
H. Aratyn and A. Das, solv-int/9710026.

\bibitem{OP}
W. Oevel and Z. Popowicz, {\it Commun. Math. Phys.\/} {\bf 139\/}, 441 (1991).

\bibitem{FMR}
J. M. Figueroa-O'Farrill, J. Mas and E. Ramos,
{\it Rev. Math. Phys.\/} {\bf 3\/}, 479 (1991).

\bibitem{Liu}
Q. P. Liu, {\it Lett. Math. Phys.\/} {\bf 35}, 115 (1995).

\bibitem{LM}
Q. P. Liu and M. Ma\~nas, {\it Phys. Lett.\/} {\bf B394}, 337 (1997); 
{\it Phys. Lett.\/} {\bf 396\/} 133 (1997); solv-int/9711002.

\bibitem{ST}
J. C. Shaw and M. H. Tu,  solv-int/9710005; solv-int/9712009.

\bibitem{ANP2}
H. Aratyn, E. Nissimov and  S. Pacheva, solv-int/9801021.

\end{thebibliography}
\end{document}